\documentstyle[12pt]{article}
\begin{document}

\rightline{IMPERIAL/TP/96-97/09}

\

\vskip 2 truecm
\

\begin{center}
{ON PHASE ORDERING BEHIND THE PROPAGATING FRONT OF A SECOND-ORDER
TRANSITION. }
\

\vskip 2 truecm
{ T.W.B. Kibble\\
Blackett Laboratory, Imperial College,\\ London SW7
2BZ, United Kingdom\\ \
\\ and \\ \
\\G.E. Volovik\\
Low Temperature Laboratory, Helsinki
University of Technology,\\
 02150 Espoo, Finland\\
and\\
L.D. Landau Institute for Theoretical Physics,\\
Kosygin Str. 2, 117940 Moscow, Russia
}
\end{center}

\begin{abstract} In a real system the heating is nonuniform and a
second-order phase transition into a broken symmetry phase occurs
by propagation of the temperature front. Two parameters, the
cooling rate $\tau_Q$ and the velocity $v_T$ of the transition
front, determine the nucleation of topological defects. Depending
on the relation of these parameters two regimes are found: in the
regime of fast propagation defects are created according to the Zurek
scenario for the homogeneous case, while in the slow propagation
regime vortex formation is suppressed.
\end{abstract}


\vfill\eject

\centerline{\bf 1. Introduction.}

A common issue of particular interest both in cosmology and
in condensed matter physics is an estimation of the initial defect
density produced during  a phase transition into a broken symmetry
state \cite{Kibble,Zurek}.  In cosmology this is an important
issue so far as the later transitions are concerned --- for
example in extended models where there are stable defects produced
at the electroweak transition.  It is probably not of much
practical importance for the GUT transition (if our ideas about
scaling are correct) because in that case so much time has elapsed
that no trace of the initial conditions remains
\cite{Vilenkin,CosmicStrings}. The same thing is true no doubt for
some condensed matter systems, but it would be good to identify
some situations in which predictions about the initial density
(not just the later scaling value) could be tested.  We really
need to find cases (i) in which the transition is second-order,
(ii) in which it is of first order and goes by bubble nucleation,
and (iii) in which it goes by the `false vacuum' becoming totally
unstable, which has been called spinodal decomposition. All these
cases correspond to different scenarios of the phase ordering
below the phase transition, and the creation and evolution of the
defects prior to the final establishment of long-range order
(phase coherence, or generally the coherence of the Goldstone
variables).

The great thing about superfluid helium-3 is that it allows such
a wide range of possibilities. Most of these cases, or possibly
all of them, may be represented in different regions of the
parameter space: the normal $^3$He to $^3$He-B or normal $^3$He to
$^3$He-A transition is second order, while the A-to-B transition
is first order and proceeds in different ways depending on the
pressure (and magnetic field) and the extent of supercooling
\cite{VolWol}.  Also of course, it adds another example to the list of possible
systems, including nematics \cite{CosmologyLiqCrystal}, helium-4
\cite{Cosmology4He} and high- and low-temperature superconductors,
but an example which is in important ways more relevant since the
fermion-boson interaction in $^3$He shares many properties of the
quantum field theory in high energy physics \cite{Exotic}. This is
why it is so valuable as an analogue of the early universe.

One of the most interesting experiments, which can shed light on
the phase ordering is the `mini big bang' produced by thermal
neutrons \cite{BigBangNature1,BigBangNature2}. There the
exothermic nuclear reaction $n+ ^3_2$He $\to p+ ^3_1$H + 0.76 MeV
produces a local Big Bang in $^3$He-B --- a region of high
temperature, $T>T_c$, where the symmetry is restored. The
subsequent cooling of this region back through the second-order
superfluid transition results in the creation of a network of
vortex lines. These seeds of vortex lines are grown by the
applied rotation and measured with the NMR technique
\cite{BigBangNature1}.

Theoretically the formation of defects in a second-order
transition into a broken symmetry state,  discussed in
\cite{Kibble,Zurek}, corresponds to a homogeneous phase transition.
In real experiments the heating is never uniform. For example,
in the neutron experiments the heating above $T_c$ by the neutron
and the subsequent cooling occur locally in a region of about
$10~\mu$m size. This results in a temperature gradient and thus
in a propagating boundary of the second-order transition. The
temperature gradient is present also in other experiments where
defect formation is influenced by the details of the cooling
through $T_c$; see e.g.\ \cite{A-vorticesVsCoolingRate} where the
type of $^3$He-A vortices created depends on the cooling rate.
One might think that the phase of the order parameter would be
determined by the ordered state behind the propagating boundary and
that vortex formation would be suppressed in this geometry. On the
other hand, in the limit of very rapid motion of the boundary
one comes to the situation of an instantaneous transition and the
Kibble-Zurek mechanism should be restored. Here we discuss the
criterion on the propagation velocity of a second-order phase
transition which separates the two regimes. We shall illustrate
it with the example of the experimentally studied normal $^3$He to
$^3$He-A and normal $^3$He to $^3$He-B transitions.

\bigskip
\centerline{\bf 2. Phase ordering in a spatially homogeneous
transition}

A nonequilibrium phase transition into the state in which the
symmetry $U(1)$ is broken leads to the formation of an infinite
cluster of the topological defects -- vortices or strings
\cite{Kibble}.  As a result, when the temperature crosses $T_c$ two
length scales appear. One of them is the  conventional
coherence length, which diverges at $T_c$
$$\xi \approx\xi_0\left(
1-{{T }\over{T_c}}\right)^{-\nu}~~.\eqno(2.1)$$
(For superfluid
$^3$He, where the Ginzburg-Landau theory is valid, one has
$\nu=1/2$). Another one is the mean distance $\tilde\xi$ between
the vortices in the infinite cluster; it defines the scale within
which the phase of the order parameter is correlated. This scale
divereges with time when the vortex cluster decays. We are
interested in the estimation of the initial density of the defects,
$\tilde\xi_{\rm initial}$, i.e., at the moment when these two scales
become well defined.

According to Zurek \cite{Zurek},  $\tilde\xi_{\rm
initial}$ is determined by the cooling rate $\tau_Q$, which
characterizes the time dependence of the temperature in the vicinity of the
phase transition:
$$~~ \epsilon(t)\equiv 1-{{T(t)}\over{T_c}}\approx
{{t}\over{\tau_Q}}~~.\eqno(2.2)$$
Well defined vortices
are formed at the time when the regions within the
Ginzburg-Landau coherence length become causally connected.
Causal connection is established by the propagation of the
order parameter. In the case of superfluid $^3$He the propagation
velocity of the order parameter can be estimated as the
velocity of spin waves which just represent the propagating
oscillations of some components of the order parameter. Thus one
has that the corresponding velocity also depends on time and is
given by
$$c(t)\approx c_0\epsilon^{1/2}(t)~~,\eqno(2.3)$$
where $c_0$ is of order of the Fermi velocity $v_F\sim 10^4~$cm.

At the moment $t\approx t_{coh}$ when
$$\xi(t_{coh})\approx\int_0^{t_{coh}} c(t')dt'~~,\eqno(2.4)$$
the regions within the coherence length $\xi(t_{coh})$ are already
connected, while outside they are causally disconnected. So the
phase of the order parameter is well defined within the coherence
length, but the phases in regions outside $\xi$ do not match each
other. This corresponds to well defined vortices with the
separation $\tilde\xi_{\rm
initial}=\xi(t_{coh})$. The time $t_{coh}$ after the transition
when this happens is
$$t_{coh}\approx\sqrt{\tau_0\tau_Q}~~,~~\tau_0\approx{{\xi_0}\over
{v_F}}\approx{{\hbar}\over
{\Delta_0}}\sim 10^{-9}~{\rm s}~~, \eqno(2.5)$$
and thus the initial separation between the vortices in the
infinite cluster is
$$\tilde\xi_{\rm
initial}=\xi(t_{coh})\approx\xi_0({{\tau_Q}\over
{\tau_0}})^{1/4}~~.\eqno(2.6)$$
For the more general case, when the phase transtion is not
necessarily described by the Ginzburg-Landau theory, one has
$\tilde\xi_{\rm initial}\sim \xi_0( \tau_Q/\tau_0)^{\alpha}$.

The further development of the infinite cluster, which leads to its
final elimination, has been the subject of the intensive
investigations in phase ordering kinetics (see e.g. the review
\cite{Bray}).

In the neutron experiments in $^3$He-B the estimation for $\tau_Q
\sim R_b^2/D$, where $R_b$ is the size of the bubble and $D$ is
the diffusion constant near $T_c$. This gives $\tau_Q \sim
10~\mu$s, and $\tilde\xi_{\rm
initial}\sim 10^{-4}$ cm. In the typical A-phase
experiments the cooling of the A-phase through the transition from
the normal liquid N occurs during $\tau_Q\sim 10^3-10^4$ s
\cite{A-vorticesVsCoolingRate}, which gives
$\tilde\xi_{\rm
initial}\sim 10^{-2}$ cm.

\bigskip
\centerline{\bf 3. The two regimes for a moving front}

In the presence of a temperature gradient two parameters
characterize the noneqilibrium phase transition. In addition to
$\tau_Q$, which determines the time scale of the temperature
change, one has now the characteristic length scale of the
temperature
$${1\over \lambda}\approx{|\nabla T|\over T_c}~~.\eqno(3.1)$$
Combining with $\tau_Q$  this gives the velocity $v_T$ of the
propagating temperature front:
$$v_T\approx{\lambda\over \tau_Q}~~,\eqno(3.2)$$
which is thus the velocity of the propagating second-order
transition.

The homogeneous result Eq.~(2.6) is obtained when the velocity of
the front is large compared to $c(t_{coh})$, so the causality
argument does work. This gives the estmation for the critical
value of the velocity
$$v_{Tc}\approx{ \xi(t_{coh})\over
t_{coh}}\approx c_0({{\tau_0}\over {\tau_Q}})^{1/4} ~~.
\eqno(3.3)$$
If $ v_T < v_{Tc}$, the slowly moving front dictates the phase of
the order parameter and  formation of vortices should be
significantly suppressed compared to the case of a rapidly moving
front. In
$^3$He-B neutron experiments one has
$\lambda \sim R_b\sim 10~\mu$m (where $R_b$ is the maximum radius of
the bubble of normal fluid above $T_c$) and
$v_T\sim 10^3$~cm/s. This is comparable with
$c_0(\tau_0/\tau_Q)^{1/4} \sim 10^3$~cm/s so vortex nucleation is not
suppressed by the moving interface in this experiment.

In the A-phase experiments the typical $\lambda\sim 10^2$ cm and
$v_T\approx\lambda\tau_Q\sim 10^{-2}-10^{-1}$ cm/s, which is much
less than $c(t_{coh}) \sim 10$ cm/s. Thus at this low $v_T$
Eq.~(2.6) does not hold since the phase correlation across the
front occurs faster than due to the temporal change of $T$. So
the creation of vortices is markedly suppressed. However even in
this regime formation of defects has been experimentally
observed, the formation of planar solitons in $^3$He-A
\cite{SolitonNucleation}.

\bigskip
\centerline{\bf 4. Phase ordering behind a slowly moving front.}

One may expect that in the whole range of $v_T$ the initial density
of vortices immediately behind the front  is determined by the
general scaling law:
$${\tilde\xi_{\rm initial}(v_T,\tau_Q)\over \xi_0}= x^{\alpha}
F({x^{\beta} y})~~,~~x={{\tau_Q}\over {\tau_0}}~~,~~y={v_T\over c_0}
  ~~ ,\eqno(4.1)$$
where $F(u)$ is some function, which has different asymptotes in
the two regimes discussed above. From the previous sections it follows
that for $^3$He one has $\alpha=\beta=1/4$. The regime of fast
propagation of the temperature front corresponds to the
asymptote $F(u) \rightarrow 1$ when its argument
$u\gg 1$. Let us find the asymptote in the regime of slow
propagation, i.e. $F(u\ll 1)$.

Since the propagation velocity of the order parameter $c(t)$ slows
down very near the transition, in some neighbourhood of the front,
in a layer of thickness $\Delta z\approx v_T \Delta t$, the
situation becomes `homogeneous'. Here $\Delta t$ is obtained from
$$v_T\Delta t\approx\int_0^{\Delta t}c(t)dt\approx c_0{(\Delta
t)^{3/2}\over
\tau_Q^{1/2}}~~, \eqno(4.2)$$
which gives
$$ \Delta t\approx \tau_Q {v_T^{2}\over
c_0^{2}}~~. \eqno(4.3)$$
Outside this layer the condensate phase is already fixed due to
the phase correlations with the low temperature regions, which are
transferred by order parameter waves propagating along the vertical
axis. So the only source of the mismatch of the phase originates
within this thin layer, which means that the magnitude  of the
coherence length $\xi(T)$ within the layer determines the
initial distance between the vortices as a function of $v_T$ at slow
transition:
$$\tilde\xi_{\rm initial}(v_T)\approx\xi(t\approx\Delta
t)\approx\xi_0{c_0\over v_T} ~~,~~{v_T\over  c_0}<
({{\tau_0}\over {\tau_Q}})^{1/4}~~ .\eqno(4.4)$$
Thus in the limit $u\ll 1$, we find $F(u)\sim 1/u$, so the initial
length scale is essentially larger than in the case of a rapidly
propagating front, where:
$$\tilde\xi_{\rm initial}(v_T)\approx \xi_0({{\tau_Q}\over
{\tau_0}})^{1/4} ~~,~~{v_T\over  c_0}>  ({{\tau_0}\over
{\tau_Q}})^{1/4}~~ .\eqno(4.5)$$

Estimation of $\tilde\xi_{\rm initial}$ for a slowly
propagating front in the A-phase gives $\tilde\xi_{\rm initial}
\sim 0.01-0.1$ cm. This is significantly smaller than the size of
the cell, $\sim 1$ cm, and thus well explains the appearance of
solitons during the cooling into the A-phase in
\cite{SolitonNucleation}.

On the other hand this is of order of the intervortex distance in
the rotating cryostat and thus can influence the vortex texture
which appears in the rotating cryostat when the superfluid
transition occurs under rotation
\cite{A-vorticesVsCoolingRate}. In a field of 10 mT two types of
vortices are competing in the equilibrium rotating state: singular
one-quantum vortices with the core radius
$r_{core}\sim \xi(T)$ and continuous two-quantum vortices
(textures) with $r_{core}\sim \xi_D$, the dipole length
$\xi_D\sim 10^{-3}$ cm.  At low rotation velocities, $\Omega < 2$
rad/s (or in general at low vortex density), the array of singular
one-quantum vortices has less energy. The experimental evidence is
that well below 1 rad/s singular vortices are created after
the superfluid transition under rotation, and no dependence on
the cooling rate was observed. However, the experiment at
higher velocity, $\Omega \approx1.4$ rad/s, showed such dependence:
at slow transition the equilibrium (singular) vortices dominate
in the cell --- their fraction is
$n_s/(n_s+2n_c)\sim $ 0.8 --- while at fast transition the fraction
of continuous vortices $2n_c/(n_s+2n_c)$  sharply increases from
0.2 up to 0.8. The change occurs in a jump-like manner at
$\partial_t T\sim 6 \mu$K/min which corresponds to
$\tau_Q\sim 3\cdot 10^4$ s. One may expect that this is related
to the phase ordering process, which leads to an initial vortex
density larger than the equilibrium one. The further
relaxation of the intial network towards equilibrium may
discriminate between textures and singular vortices, which have
significantly different scales.

\bigskip
\centerline{\bf 5. Conclusion.}

In a spatially inhomogeneous phase transition vortex
formation depends on the velocity $v_T$ of the propagating front
of the second-order phase transition. There is a critical value
of the front velocity, $ v_{Tc} \approx c_0 (\tau_/\tau_Q)^{1/4}$,
which separates two regimes. For a rapidly propagating front, with
$v_T > v_{Tc}$, vortex formation is the same as in a
spatially homogeneous phase transition. For a slowly propagating
front, with $v_T < v_{Tc}$, vortex formation becomes less
favorable whith decreasing $v_T$.

We thanks M. Krusius and \"U. Parts for numerous discussions  of
the $^3$He experiments. This work was supported through the ROTA
co-operation plan of the Finnish  Academy and the Russian Academy
of Sciences, by the Russian Foundation for Fundamental Sciences,
Grant No. 96-02-16072, and  was carried out under the EU Human
Capital and Mobility Programme (contract numbers CHGE-CT94-0069 and
CHRX-CT94-0423).

\end{document}